\documentclass[10pt,conference]{IEEEtran}
\IEEEoverridecommandlockouts
\usepackage{amsmath,amssymb,amsfonts}
\usepackage{graphicx}
\usepackage{textcomp}
\usepackage{listings}
\usepackage{xcolor}
\usepackage{xspace}
\definecolor{codegreen}{rgb}{0,0.6,0}
\definecolor{codegray}{rgb}{0.5,0.5,0.5}
\definecolor{codepurple}{rgb}{0.58,0,0.82}
\definecolor{backcolour}{rgb}{1,1,1}

\lstdefinestyle{mystyle}{
    backgroundcolor=\color{backcolour},   
    commentstyle=\color{codegreen},
    keywordstyle=\color{magenta},
    stringstyle=\color{codepurple},
    basicstyle=\ttfamily\tiny,
    breakatwhitespace=false,         
    breaklines=true,                 
    captionpos=b,                    
    keepspaces=true,                 
    showspaces=false,                
    showstringspaces=false,
    showtabs=false,                  
    tabsize=2,
    emphstyle=\textbf, 
    emph={Using, Decorator, Comments, (a), (b), (c), Hardcoded, Warnings}
}

\usepackage[linesnumbered,ruled,vlined]{algorithm2e}
\usepackage{algpseudocode}
\algnewcommand{\algorithmicand}{\textbf{ and }}
\usepackage{pifont}
\usepackage{program}
\usepackage{xcolor}
\usepackage{url}
\usepackage{hyperref}
\usepackage{makecell}

\SetCommentSty{mycommfont}
\def\BibTeX{{\rm B\kern-.05em{\sc i\kern-.025em b}\kern-.08em
    T\kern-.1667em\lower.7ex\hbox{E}\kern-.125emX}}

\begin{document}

\title{API\textit{Scanner} - Towards Automated Detection of Deprecated APIs in Python Libraries}

\author{\IEEEauthorblockA{Aparna Vadlamani, Rishitha Kalicheti, Sridhar Chimalakonda}
\IEEEauthorblockA{\textit{Research in Intelligent Software \& Human Analytics (RISHA) Lab}\\
\textit{Dept. of Computer Science \& Engineering} \\
\textit{Indian Institute of Technology Tirupati}\\
\textit{Tirupati, India}\\
\textit{\{cs17b005, cs17b014, ch}\}@iittp.ac.in}
}

\maketitle
\begin{abstract}
Python libraries are widely used for machine learning and scientific computing tasks today. APIs in Python libraries are deprecated due to feature enhancements and bug fixes in the same way as in other languages. These deprecated APIs are discouraged from being used in further software development. Manually detecting and replacing deprecated APIs is a tedious and time-consuming task due to the large number of API calls used in the projects. Moreover, the lack of proper documentation for these deprecated APIs makes the task challenging. To address this challenge, we propose an algorithm and a tool API\textit{Scanner} that automatically detects deprecated APIs in Python libraries. This algorithm parses the source code of the libraries using abstract syntax tree (ASTs) and identifies the deprecated APIs via \textit{decorator}, \textit{hard-coded warning} or \textit{comments}. API\textit{Scanner} is a Visual Studio Code Extension that highlights and warns the developer on the use of deprecated API elements while writing the source code. The tool can help developers to avoid using deprecated API elements without the execution of code. We tested our algorithm and tool on six popular Python libraries, which detected 838 of 871 deprecated API elements. Demo of API\textit{Scanner}: \url{https://youtu.be/1hy_ugf-iek}. Documentation, tool, and source code can be found here: \url{https://rishitha957.github.io/APIScanner}.
\end{abstract}

\begin{IEEEkeywords}
Deprecated APIs, Python Libraries, API Evolution, Visual Studio Code Extension
\end{IEEEkeywords}

\section{Introduction}
Python is one of the popular dynamic programming language that has gained immense popularity due to its extensive collection of libraries, including popular modules for machine learning and scientific computing \footnote{\url{https://www.tiobe.com/tiobe-index/}}. Due to reasons such as feature improvements and bug repairs, python libraries are frequently updated. Most API changes include moving \textit{methods} or \textit{fields} around and \textit{renaming} or \textit{changing} method signatures  \cite{refactor}. These changes may induce compatibility issues in client projects \cite{zhang2020Python}. It is recommended to follow the \textit{deprecate-replace-remove} cycle to enable developers to adapt to these changes smoothly\cite{li2020cda}. In this process, APIs that are no longer supported are first labeled as deprecated, and then the deprecated APIs are replaced with their substitution messages to help developers transition from deprecated APIs to new ones \cite{replacement}. The deprecated APIs are gradually removed from the library in future releases. Unfortunately, this process is not always followed, as discovered by several studies \cite{smalltalk, pharo}, making it difficult for both library maintainers and developers. Ko et al. have analyzed the quality of documentation for resolving deprecated APIs \cite{documentquality}. Researchers have proposed techniques to automatically update deprecated APIs \cite{haryono2020automatic, automaticupdate}. However, most of them are for static programming languages such as \textit{Java}, \textit{C\#} and \textit{Android SDKs}. Python being a typical dynamic programming language, exhibits different API evolution patterns compared to \textit{Java} \cite{zhang2020Python}. Hence it motivates the need for new techniques and tools to detect deprecated APIs.

Deprecated APIs in Python libraries are mainly declared by \textit{decorator}, \emph{hard-coded warning}, and \textit{comments} \cite{wang2020exploring}. Nevertheless, it was discovered that library maintainers use varied and multiple strategies for API deprecation, leading to inconsistency in the implementation of libraries as well as their automated detection \cite{wang2020exploring}. In addition, nearly one-third of the deprecated APIs in Python is not included in the official library documentation, making it hard for developers using libraries to limit the use of deprecated APIs \cite{wang2020exploring}. 

To avoid the usage of deprecated APIs during new software development, developers should be aware of deprecating APIs in the project, motivating the need for this research. Hence, given the rise in popularity of Python and the number of deprecated APIs used in Python projects, we propose a novel algorithm that uses the source code of the Python libraries to get a list of deprecated APIs. This list is further used to detect deprecated APIs in Python projects. This paper contributes (i) an algorithm for deprecated API detection and (ii) a Visual Studio Code extension, API\textit{Scanner}\footnote{\url{https://marketplace.visualstudio.com/items?itemName=Rishitha.apiscanner}}. We believe that API\textit{Scanner} might assist developers to detect deprecated APIs and help them avoid searching through API documentation or on forums such as Stack Overflow. As a preliminary evaluation, we tested our algorithm and tool on six popular Python libraries \cite{jupyter} that are commonly used in data analytics, machine learning, and scientific computing. The initial results are promising with 90\% API deprecation detection, with potential for application beyond these libraries. 

\section{Approach}
\begin{figure}
    \centering
    \includegraphics[width=\linewidth,height=\textheight,keepaspectratio]{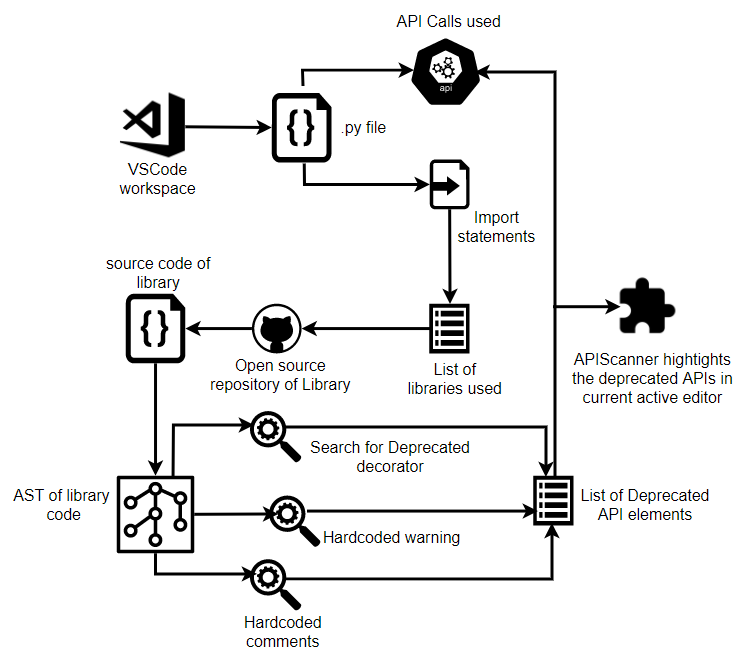}
    \caption{Approach for Detecting Deprecated API Elements in Python Libraries}
    \label{fig:my_label}
\end{figure}

\begin{figure}
    \centering
    \includegraphics[width=\linewidth,height=\textheight,keepaspectratio]{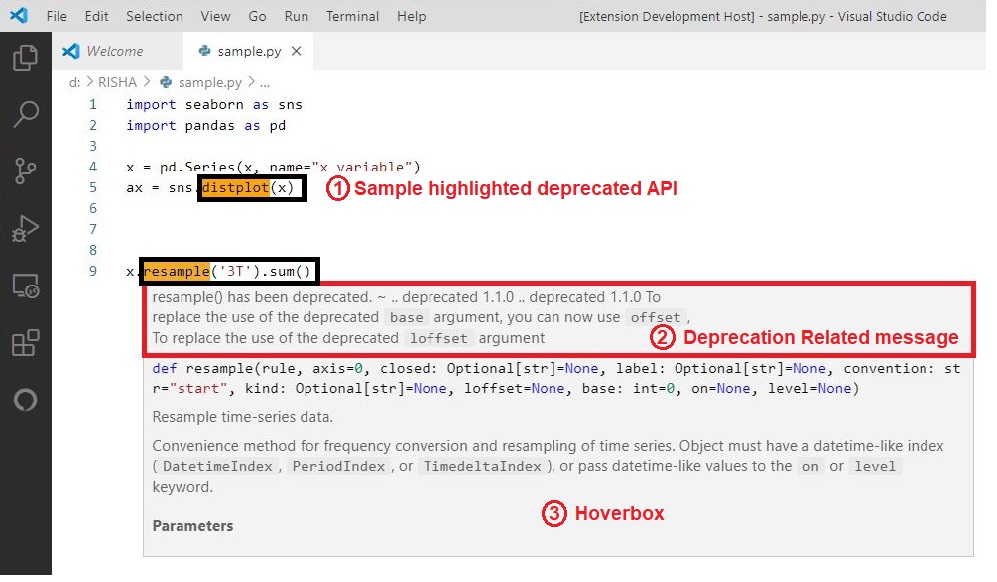}
    \caption{Snapshot of API\textit{Scanner}. The black boxes indicate deprecated APIs highlighted by API\textit{Scanner}. The red box indicates the message shown by API\textit{Scanner} on hovering over the highlighted deprecated APIs.}
    \label{fig:sampleui}
\end{figure}

Wang et al. \cite{wang2020exploring} investigated that inconsistency in the adopted deprecation strategies makes it a harder task to use automated approaches for managing deprecated APIs and their documentation. In this paper, we propose an approach (as shown in Fig. \ref{fig:my_label}) to automatically detect deprecated APIs in Python libraries and alert developers during API usage in software development. Firstly, we identify the libraries used in the client code from \textit{import} statements. We build an abstract syntax tree (AST) to parse the source code to detect the patterns. The proposed Algorithm \ref{algo:approach} is then applied on the ASTs to retrieve a list of deprecated APIs in those libraries. Based on this list, API\textit{Scanner} parses each line of code in the editor, highlights the deprecated elements in the editor. On hovering, the tool also displays a message informing the developer that some element(s) of this API call has been deprecated (as shown in Fig. \ref{fig:sampleui}). We developed API\emph{Scanner} as a Visual Studio Code extension as it supports both Python scripts and \textit{jupyter notebooks}\footnote{https://jupyter.org/}.  


\begin{lstlisting}[caption={Examples of methods of deprecation strategies adopted in Python libraries which are deprecated through a) \textit{decorator}, b) \textit{comments} c) \textit{hard-coded warning}},label={lst:label1}, language=Python, captionpos=b]
(a)  Using Decorator: in Matplotlib
    @_api.deprecated("3.3", alternative="Glue('fil')")
    class Fil(Glue):
        def __init__(self):
            super().__init__('fil')
(b) Using Comments: in Sklearn
    class GradientBoostingClassifier(args):
        """
        ..criterion : {'friedman_mse', 'mse', 'mae'}..
        .. deprecated:: 0.24 `criterion='mae'` is deprecated and will be removed in version 0.26. Use `criterion='friedman_mse'` or `'mse'` instead, as trees    should use a least-square criterion in Gradient Boosting
        """
(c) Using Hardcoded Warnings: in Pandas
    class Series(args):
        def __init__(self,args):
            if dtype is None:
                warnings.warn("The default dtype for empty Series will be 'object' instead of 'float64' in a future version",DeprecationWarning,stacklevel=2)
\end{lstlisting}


\subsection{Detecting Deprecated API Elements through Source Code}
We parse the source code of the library to generate an AST and denote it as $P_{AST}$. Examples of Python APIs deprecated by \textit{decorator}, \textit{hard-coded warnings}, and \textit{comments} are shown in listing \ref{lst:label1}. Structure of AST helps to realize the relationship between class declaration and function definition with \textit{decorator}, \textit{hard-coded warnings}, and \textit{comments}. We traverse through each node $N_{AST}$ in the AST and generate $P_{AST}$ using Depth-First Search (cf. Line-2). Whenever we encounter a class definition node, we extract the \textit{doc-string} of that particular class. If the \textit{doc-string} contains the deprecate keyword (such as (b) in Listing \ref{lst:label1}), we generate the Fully Qualified API name of the class by appending the class name to the directory path. We also append the deprecation message to $L_D$ (cf. Line-13) along with a list of decorators associated with the class. If there is a deprecated decorator (such as (a) in Listing \ref{lst:label1}) in the extracted list, we add the fully qualified name of the class and any description provided to list $L_D$ (cf. Line-16). Similarly, when we encounter the function definition node, we extract the list of decorators associated with it. If there is a deprecated decorator in the extracted list, we add a fully qualified name of the function to list $L_D$ (cf. Line-6). For each function call node in $N_{AST}$ (cf. Line-7), we verify if \textit{DeprecationWarning} or \textit{FutureWarning} are passed as arguments (such as (c) in Listing 1) and add its fully qualified name to list $L_D$, which is the final generated list of deprecated API elements.  

\begin{algorithm}
    \DontPrintSemicolon
    \scriptsize
    \SetKwInput{KwInput}{Input}
    \SetKwInOut{KwOutput}{Output}
    \KwInput{$P$, Python Library Code}
    \KwOutput{$L_D$, List of Deprecated API Elements}
    \BlankLine
    \SetKwFunction{FMain}{Detect\_Deprecated\_API}
    \SetKwProg{Fn}{Function}{:}{}
      \Fn{\FMain{}}{
            $L_D \gets \{\}$ \;
            \tcc{parseCode returns Abstract syntax tree of given code input}
            $P_{AST} \gets parseCode(P)$ \;
            \tcc{Traverse each node in  $P_{AST}$ using BFS}
            \For{$N_{AST} \in \mathcal P_{AST} $}{
                \If{isFunctionDefNode($N_{AST}$)}
                {
                    $D$ = $N_{AST}.Decorators$\;
                    \If{isDeprecatedDecorator(D)}{
                        $L_D$.add(getFullyQualifiedName($N_{AST}.Name$))
                    }
                    \tcc{Traverse each Node in  $N_{AST}$}
                    \For{$Node \in \mathcal N_{AST} $}{
                        \If{isFunctionCallNode($Node$) \algorithmicand isDeprecationWarning($Node$)}{
                            $L_D$.add(getFullyQualifiedName($N_{AST}.Name$))
                        }   
                    }
                }
                \ElseIf{isClassDefNode($N_{AST})$}
                {
                	$doc_{str}$ = $N_{AST}$.Docstring\;
                    \If{$doc_{str}$.hasDeprecationKeyword()}{
                        $L_D$.add(getFullyQualifiedName($N_{AST}.Name$))\;
                    }
                    $D$ = $N_{AST}.Decorators$\;
                    \If{isDeprecatedDecorator(D)}{
                        $L_D$.add(getFullyQualifiedName($N_{AST}.Name$))\;
                    }
                }
            }
            \KwRet $L_D$\;
      }
     \caption{Detecting Deprecated API Elements in Python Libraries}
     \label{algo:approach}
\end{algorithm}

\section{Evaluation}
\subsection{Libraries Selection}
To evaluate our approach, we applied it on six popular third-party Python libraries that were identified by Pimentel et al \cite{jupyter}. However, this approach is not limited to the selected libraries and could be applied to other Python libraries as well.

\begin{itemize}
    \item \textit{NumPy}: Array programming library \cite{harris2020array}.
    \item \textit{Matplotlib}: A 2D graphics environment \cite{Hunter:2007}.
    \item \textit{Pandas}: Data analysis and manipulation tool \cite{pandas}.
    \item \textit{Scikit-learn}: Machine learning library for Python  \cite{scikit-learn}. 
    \item \textit{Scipy}: Library for scientific and technical computing \cite{virtanen2020scipy}.
    \item \textit{Seaborn}: Data visualization based on \textit{matplotlib} \cite{waskom2020seaborn}.
\end{itemize}

\subsection{Results}
Table \ref{tab:results} summarizes the total number of deprecated API elements detected by the Algorithm \ref{algo:approach} and the total number of deprecated API elements found in the source code of the Python libraries. We manually counted the number of deprecated API elements present in the source code of the libraries. From Table \ref{tab:results}, we can observe that the algorithm has detected more than 90\% of the deprecated APIs. In the case of \textit{Matplotlib}, only 65\% of the deprecated APIs could be detected since \textit{Matplotlib} deprecates many of its parameters using a custom warning function which does not have any parameters indicating if it is a \textit{DeprecationWarning} or not. In such cases, the proposed algorithm could not detect the deprecated API elements. 

In the case of \textit{Scikit-learn}, \textit{Numpy} and \textit{Pandas}, some of the functions that are used to deprecate parameters or parameter values or deprecation warnings induced by other libraries are also captured. Hence, the number of deprecated API elements detected by the algorithm is higher than the actual number of deprecated APIs. Whereas in the case of \textit{Scipy} and \textit{Seaborn}, some of the parameters are deprecated without using any of the three deprecation strategies, which could not be detected by the algorithm. Hence, the number of deprecated API elements detected by the algorithm for \textit{Scipy} and \textit{Seaborn} are lower than the actual number of deprecated APIs.
\begin{table}

\begin{tabular}{ |p{1.5cm}|p{1cm}|p{2.25cm}|p{2.25cm}|  }
 \hline
\textbf{\textbf{Library Name}} & \textbf{LOC} & \textbf{Total No. of Deprecated API elements identified using Algorithm} \ref{algo:approach} & \textbf{Total No. of Deprecated API elements in source code}\\
 \hline
 \textit{Scikit-learn} & 388.1k & 487 & 438 \\
 \hline
 \textit{Matplotlib} & 982.5k & 169 & 254 \\
 \hline
 \textit{Numpy} & 145.6k & 39 & 36 \\
 \hline
 \textit{Pandas} & 668.9k & 66 & 59 \\
 \hline
 \textit{Scipy} & 725.62k & 46 & 49\\
 \hline
 \textit{Seaborn} & 83.7k & 31 & 35 \\
 \hline 
\end{tabular}
\caption{Evaluation of results obtained using our algorithm}
\label{tab:results}
\end{table}
\section{Limitations and Threats to Validity}
API\textit{Scanner} detects deprecated APIs through \textit{decorator}, \textit{warning} or \textit{comments}. Any other deprecated APIs that are not implemented through the above three strategies cannot be detected by the algorithm. Moreover, the algorithm finds the function or class in which a parameter is deprecated but the exact parameter deprecated may not be mentioned in the deprecation message displayed by the extension due to the inconsistent deprecation strategies adopted by the library maintainers. APIs deprecated without using the \textit{DeprecationWarning} and \textit{FutureWarning} as parameters in the warning function cannot be detected by the algorithm. APIs deprecated using single-line comments and not using the doc-strings also cannot be detected by the algorithm. Further, a major pre-requisite for our approach is the availability of source code of libraries. We can mitigate the threat due to inconsistent deprecation strategies if we can ensure that the documentation is structured and well maintained for Python libraries.  

Finally, since the results are evaluated manually, there may be human errors. Hence, we have carefully reviewed and validated some of the results using release notes to mitigate this potential threat. We plan to extend the evaluation of the tool using release notes and API documentation. 

\section{Related Work}
In the literature, several studies on deprecated APIs for different environments have been done to analyze and tackle the challenges posed by the deprecation of APIs in libraries.

Robbes et al. \cite{smalltalk, pharo} studied the reactions of developers to the deprecation and the impact of API deprecation on the \textit{Smalltalk} and \textit{Pharo} ecosystem. Ko et al. \cite{documentquality} examined 260 deprecated APIs from eight Java libraries and their documentation and observed that 61\% of deprecated APIs are offered with replacements. Similarly, Brito et al. \cite{deprecateapisreplacement} conducted a large-scale study on 661 real-world Java systems and found that replacements are provided for 64\% of the deprecated APIs. In another study \cite{replacement} conducted on \textit{Java} and \textit{C\#} projects, they have observed that an average of 66.7\% of APIs in \textit{Java} projects and 77.8\% in \textit{C\#} projects were deprecated with replacement messages. In 26 open-source Java systems over 690 versions, Zhou et al. \cite{retrospective} analysed the history of deprecated APIs and observed that deprecated API messages are not well managed by library contributors with very few deprecated APIs being listed with replacements. Li et al. \cite{li2020cda} characterized the deprecated APIs in Android Apps parsing the code of 10000 Android applications. Zhang et al. \cite{zhang2020Python} have observed a significant difference in evolution patterns of Python and \textit{Java} APIs and also identified 14 patterns in which Python APIs evolve. Wang et al. \cite{wang2020exploring} observed that library contributors do not properly handle API deprecation in Python libraries. To this end, there is a need for approaches and tools to automatically detect deprecated API elements in Python projects.  

Several approaches have been proposed in the literature for other ecosystems to migrate from deprecated APIs \cite{migratedeprecateapi, automaticupdate, haryono2020automatic}. Yaoguo Xi et al. \cite{migratedeprecateapi} proposed an approach and built a tool \textit{DAAMT} to migrate from deprecated APIs in Java to their replacements if recorded in the documentation. Fazzini et al. \cite{automaticupdate} developed a technique \textit{AppEvolve} to update API changes in Android Apps by automatically learning from examples before and after-updates. Haryono et al. \cite{haryono2020automatic} proposed an approach named \textit{CocciEvolve} that updates using only a single after-update example. However, tools that handle deprecated APIs in Python projects have not been developed, which motivated us towards the development of API\textit{Scanner}. 


\section{Conclusion and Future Work}
Considering the extensive use of deprecated APIs during software development and lack of proper documentation for deprecated APIs, we proposed an approach to automatically detect deprecated APIs in Python libraries during the development phase of the project. In this paper, we presented a novel algorithm and a tool called API\textit{Scanner} that detects deprecated APIs. The algorithm identifies the APIs deprecated via \textit{decorator}, \textit{hard-coded warning} or \textit{comments} by parsing the source code of the libraries and generated a list of deprecated APIs. API\textit{Scanner} used this list and searched for the use of deprecated APIs in the current active editor. The tool highlights deprecated APIs in the source code along with further deprecation details. API\textit{Scanner} thus aims to help developers detect deprecated APIs during the development stage and avoid searching through API documentation or forums such as Stack Overflow. Highlighting the use of deprecated APIs in the editor might help developers to address and replace them. The proposed algorithm identified 838 out of 871 API elements across six different Python libraries.  


As future work, our goal is to strengthen the tool with release-specific information and develop a better user interface (such as different colors) to indicate the severity of the deprecation. We also plan to improve the documentation of deprecated APIs through the information obtained from the algorithm. We plan to extend the tool to provide a feature to migrate from the deprecated API to its replacement. We aim to improve the tool's accuracy by extracting APIs that are deprecated using the custom deprecation strategies. Finally, we plan to conduct extensive developer studies on the usage of the approach and the tool with more libraries. 

\bibliographystyle{IEEEtran}
\bibliography{main}
\vspace{12pt}
\end{document}